\font\bbb=msbm10                                                   

\def\Z{\hbox{\bbb Z}}

\def\AAM{{\sl Advances in Appl.\ Math.}}
\def\AER{{\sl Amer.\ Econom.\ Rev.}}

\def\AnM{{\sl Ann.\ Math.}}
\def\AMM{{\sl Amer.\ Math.\ Monthly}}

\def\APSR{{\sl Amer.\ Pol.\ Sci.\ Rev.}}

\def\BS{{\sl Behav.\ Sci.}}
\def\BSTJ{{\sl Bell System Tech.\ J.}}

\def\Ec{{\sl Econometrica}}
\def\EA{{\it \'Economie Appliqu\'ee}}

\def\IJCS{{\sl InterJournal Complex Systems}}

\def\JCR{{\sl J. Conflict Resolution}}

\def\JET{{\sl J. Econom.\ Theory}}

\def\JME{{\sl J. Math.\ Econom.}}

\def\JMP{{\sl J. Math.\ Phys.}}

\def\JPE{{\sl J. Pol.\ Econom.}}

\def\MARS{{\it M\'emoires de l'Acad\'emie Royale des Sciences}}

\def\NP{{\sl Nucl.\ Phys.}}

\def\RAS{{\sl Robotics and Autonomous Systems}}
\def\RES{{\sl Rev.\ Econom.\ Stud.}}

\def\Sc{{\sl Science}}
\def\SCW{{\sl Soc.\ Choice Welf.}}

\def\TAMS{{\sl Trans.\ Amer.\ Math.\ Soc.}}

\def\dajm{\hbox{D. A. Meyer}}

\def\brosl{\hbox{B. Hasslacher}}

\def\mckelvey{\hbox{R. D. McKelvey}}

\def\hfb{\hfil\break}

\catcode`@=11
\newskip\ttglue

   \font\ninerm=cmr9    \font\eightrm=cmr8   \font\sixrm=cmr6
  \font\ninebf=cmbx9   \font\eightbf=cmbx8  \font\sixbf=cmbx6
  \font\nineit=cmti9   \font\eightit=cmti8  
  \font\ninesl=cmsl9   \font\eightsl=cmsl8  
  \font\ninemi=cmmi9   \font\eightmi=cmmi8  \font\sixmi=cmmi6

\font\bigtenbf=cmr10 scaled\magstep2 

\def\ninepoint{\def\rm{\fam0\ninerm}%
  \textfont0=\ninerm \scriptfont0=\sixrm
  \textfont1=\ninemi \scriptfont1=\sixmi
  \textfont\itfam=\nineit  \def\it{\fam\itfam\nineit}%
  \textfont\slfam=\ninesl  \def\sl{\fam\slfam\ninesl}%
  \textfont\bffam=\ninebf  \scriptfont\bffam=\sixbf
    \def\bf{\fam\bffam\ninebf}%
  \tt \ttglue=.5em plus.25em minus.15em
  \normalbaselineskip=11pt
  \setbox\strutbox=\hbox{\vrule height8pt depth3pt width0pt}%
  \normalbaselines\rm}

\def\eightpoint{\def\rm{\fam0\eightrm}%
  \textfont0=\eightrm \scriptfont0=\sixrm
  \textfont1=\eightmi \scriptfont1=\sixmi
  \textfont\itfam=\eightit  \def\it{\fam\itfam\eightit}%
  \textfont\slfam=\eightsl  \def\sl{\fam\slfam\eightsl}%
  \textfont\bffam=\eightbf  \scriptfont\bffam=\sixbf
    \def\bf{\fam\bffam\eightbf}%
  \tt \ttglue=.5em plus.25em minus.15em
  \normalbaselineskip=9pt
  \setbox\strutbox=\hbox{\vrule height7pt depth2pt width0pt}%
  \normalbaselines\rm}

\def\sfootnote#1{\edef\@sf{\spacefactor\the\spacefactor}#1\@sf
      \insert\footins\bgroup\eightpoint
      \interlinepenalty100 \let\par=\endgraf
        \leftskip=0pt \rightskip=0pt
        \splittopskip=10pt plus 1pt minus 1pt \floatingpenalty=20000
        \parskip=0pt\smallskip\item{#1}\bgroup\strut\aftergroup\@foot\let\next}
\skip\footins=12pt plus 2pt minus 2pt
\dimen\footins=30pc

\def\ie{{\it i.e.}}
\def\eg{{\it e.g.}}

\def\Proof{{\sl Proof}}

\def\Proposition{P{\eightpoint ROPOSITION}}

\def\endproof{\vrule height6pt width6pt depth0pt}
\def\@versim#1#2{\lower.5pt\vbox{\baselineskip0pt \lineskip-.5pt
    \ialign{$\m@th#1\hfil##\hfil$\crcr#2\crcr\sim\crcr}}}
\def\gsim{\mathrel{\mathpalette\@versim\>}}

\magnification=1200
\input epsf.tex

\dimen0=\hsize \divide\dimen0 by 13 \dimendef\chasm=0
\dimen1=\hsize \advance\dimen1 by -\chasm \dimendef\usewidth=1
\dimen2=\usewidth \divide\dimen2 by 2 \dimendef\halfwidth=2
\dimen3=\usewidth \divide\dimen3 by 3 \dimendef\thirdwidth=3
\dimen4=\hsize \advance\dimen4 by -\halfwidth \dimendef\secondstart=4
\dimen5=\halfwidth \advance\dimen5 by -10pt \dimendef\indenthalfwidth=5
\dimen6=\thirdwidth \multiply\dimen6 by 2 \dimendef\twothirdswidth=6
\dimen7=\twothirdswidth \divide\dimen7 by 4 \dimendef\qttw=7
\dimen8=\qttw \divide\dimen8 by 4 \dimendef\qqttw=8
\dimen9=\qqttw \divide\dimen9 by 4 \dimendef\qqqttw=9

\parskip=0pt\parindent=0pt

\line{\hfil June 1997}
\line{\hfil{\it revised\/} June 1998}
\line{\hfil cond-mat/9806359}
\bigskip\bigskip\bigskip
\centerline{\bf\bigtenbf STATISTICAL MECHANICS OF VOTING}
\vfill
\centerline{\bf David A. Meyer$^*$ and Thad A. Brown$^{\dagger}$}
\bigskip
\centerline{\sl Center for Social Computation and 
                Institute for Physical Sciences}
\smallskip
\centerline{\sl $^*$Project in Geometry and Physics,
                Department of Mathematics}
\centerline{\sl University of California/San Diego,
                La Jolla, CA 92093-0112}
\centerline{dmeyer@chonji.ucsd.edu}
\smallskip
\centerline{\sl $^{\dagger}$Department of Political Science, 
                113 Professional Building}
\centerline{\sl University of Missouri/Columbia,
                Columbia, MO 65211}
\centerline{polstab@showme.missouri.edu}
\vfill
\centerline{ABSTRACT}
\bigskip
\noindent Decision procedures aggregating the preferences of multiple 
agents can produce cycles and hence outcomes which have been described 
heuristically as `chaotic'.  We make this description precise by 
constructing an explicit dynamical system from the agents' preferences 
and a voting rule.  The dynamics form a one dimensional statistical 
mechanics model; this suggests the use of the topological entropy to 
quantify the complexity of the system.  We formulate natural 
political/social questions about the expected complexity of a voting 
rule and degree of cohesion/diversity among agents in terms of random 
matrix models---ensembles of statistical mechanics models---and 
compute quantitative answers in some representative cases.

\bigskip\bigskip
\noindent{\sl Journal of Economic Literature\/} Classification System:
                   D71,      
                   D72,      
                   C69.      

\noindent 1996 Physics and Astronomy Classification Scheme:
                   05.45.+b, 
                   05.20.-y, 
                   89.90.+n. 

\noindent American Mathematical Society Subject Classification:
                   90A28,    
                   90A08,    
                   54H20,    
                   82B44.    

\global\setbox1=\hbox{Key Words:\enspace}
\parindent=\wd1
\item{Key Words:}  Condorcet cycle, Arrow's theorem, chaotic dynamical 
                   system, topological entropy, random matrix model, 
                   political cohesion, social diversity.

\vfill
\eject

\headline{\ninepoint\it Statistical mechanics of voting       
                                                 \hfil Meyer \& Brown}
\parskip=10pt
\parindent=20pt

The input for many mathematical models of social, political and 
economic systems includes a list of preference orders, one for each 
agent in the model.  These preferences are aggregated, respectively,
by some social welfare function, or voting rule, or market mechanism.
More than 200 years ago, however, Condorcet recognized potential 
problems with voting rules, namely that aggregation might produce 
cycles [1].  For example, suppose that there are three alternatives 
$\{a,b,c\}$ and three voters rank them in the orders $a > b > c$ (by 
which we mean $a$ is preferred to $b$ which is preferred to $c$), 
$b > c > a$, and $c > a > b$.  Given a choice between $b$ and $a$, a 
2:1 majority prefers $a$; if they are offered the opportunity to 
switch from $a$ to $c$, again a majority will vote to do so; finally, 
a majority also prefers $b$ to $c$, completing a cycle.  

While this example may seem contrived, Arrow's celebrated theorem [2]
states that among an apparently reasonable set of voting rules, the 
only ones which do not encounter peculiarities of this sort for some 
{\sl profile\/} (list of preference orders) are dictatorial, \ie, they 
depend only on the preference order of a single, specified, voter.  
Taking the example seriously then, we conceive it as describing a 
sequence of states (the successive preferred alternatives), a 
situation which is naturally modelled as a dynamical system.  A 
similar perspective was originally suggested by Saari [3]; in this 
letter, motivated in part by potential applications to autonomous
machines [4] choosing new states from a sequence of alternatives, 
rather than analyzing the situation by {\sl analogy\/} with dynamical 
systems, we construct an {\sl explicit\/} map from a profile and 
voting rule to a discrete dynamical system.

\moveright\secondstart\vtop to 0pt{\hsize=\halfwidth
$$
\epsfxsize=\halfwidth\epsfbox{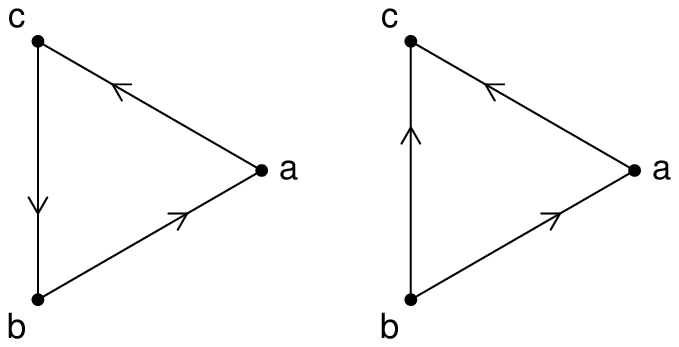}
$$
\vskip 0.25\baselineskip
\eightpoint{%
\noindent{\bf Fig.~1}.  The weak tournaments corresponding to majority 
rule on the profiles $p_1 = (a>b>c,b>c>a,c>a>b)$ and 
$p_2 = (a>b>c,c>b>a,c>a>b)$.  We suppress the edges connecting each 
vertex to itself.  Note the nontrivial cycle in the graph $f_{p_1}$
(on the left).
}}
\vskip -\baselineskip
\parshape=16
0pt \halfwidth
0pt \halfwidth
0pt \halfwidth
0pt \halfwidth
0pt \halfwidth
0pt \halfwidth
0pt \halfwidth
0pt \halfwidth
0pt \halfwidth
0pt \halfwidth
0pt \halfwidth
0pt \halfwidth
0pt \halfwidth
0pt \halfwidth
0pt \halfwidth
0pt \hsize
The usual model for a preference order is a relation, denoted $\ge$,
which is {\sl complete\/} (for all pairs of alternatives $a \ge b$ or
$b \ge a$) and {\sl transitive\/} (if $a \ge b $ and $b \ge c$ then 
$a \ge c$) [2].  When $a \ge b$ and $b \ge a$, the voter with this
preference order is {\sl indifferent\/} between $a$ and $b$; when only
$a \ge b$, say, the voter {\sl strictly\/} prefers $a$ and we write
$a > b$.  We consider aggregation formalized by maps $f$ from 
preference profiles $p$ to directed graphs $f_p$.  A directed edge 
$a \leftarrow b$ in $f_p$ indicates that for profile $p$ the map $f$ 
chooses alternative $a$ over alternative $b$.  We call $f$ a 
{\sl voting rule\/} if for all profiles $p$, $f_p$ is {\sl complete\/} 
or a {\sl weak tournament\/} [5] (for all pairs of alternatives 
$a \leftarrow b$ or $b \leftarrow a$ in $f_p$) and {\sl Pareto\/} or 
{\sl unanimous\/} (if $a \ge b$ in each preference order in $p$ then 
$a \leftarrow b$ in $f_p$).  Notice that for every alternative $x$, 
since $x \ge x$ in every preference order, $x \leftarrow x$ in $f_p$ 
for every profile and voting rule.  The weak tournament for the 
profile and majority voting rule of the example in the first paragraph 
is shown on the left in Fig.~1; we omit the edges connecting each 
vertex to itself.

We have motivated the introduction of weak tournaments by an example 
of what is essentially an amendment procedure [6], \ie, successive
pairwise votes between a new alternative and the current one.  Notice
that this definition of voting rule is actually a generalization of 
the more familiar one which requires the outcome to be a preference 
order on the set of alternatives.  Such an outcome corresponds to 
$f_p$ being transitive as well as complete and Pareto.  But Arrow's 
theorem [2], for example, says that a broader definition is necessary 
if we forbid dictatorial rules and impose the condition of 
{\sl independence of irrelevant alternatives (IIA)}---that the 
relation between $a$ and $b$ in $f_p$ depend only on the relations of 
$a$ and $b$ in the preference orders in $p$ [7].  In fact, the 
broader definition applies equally well to voters whose pairwise 
preferences are not necessarily consistent, \ie, transitive.

The directed graph which is the image of a voting rule $f$ on a 
specific profile $p$ defines a {\sl symbolic dynamical system\/}:  
Suppose the voters are presented with a sequence of alternatives---an 
agenda, by extension of the usual meaning to allow arbitrarily long 
sequences.  The results of successive pairwise votes between the new 
alternative and the current one form a sequence of symbols 
representing the chosen alternatives.  The possible sequences are 
exactly the directed paths in $f_p$, \eg, for the first example in 
Fig.~1, $baacbbaccc\ldots$\ are the first 10 symbols of an 
{\sl admissible\/} sequence/path in $f_{p_1}$.  For contrast, examine 
the second example in Fig.~1, obtained by applying the same majority 
voting rule to the profile $p_2 = (a>b>c,c>b>a,c>a>b)$.  An admissible 
sequence/path in $f_{p_2}$ can start the same way:  
$baaccccccc\ldots$, but once alternative $c$ is chosen, no other 
alternative can beat it; the sequence terminates with a string of 
$c\,$s.  It is clear that the space of admissible paths on $f_p$ 
completely characterizes a profile/voting rule pair.  This space, 
together with the {\sl shift map\/} (deletion of the first symbol of a 
sequence) forms the promised dynamical system---a {\sl (one-sided) 
subshift of finite type\/} [8].

In Fig.~1, the first set of admissible sequences seems more 
interesting/complex than the second.  To quantify this perception we
enumerate the admissible sequences which are periodic with period $N$:
Define the {\sl transition matrix\/} $F_p$ by $(F_p)_{ab} = 1$ if
$a \leftarrow b$ in $f_p$ and $(F_p)_{ab} = 0$ otherwise.  Then the
number of $N$-periodic sequences is the trace of $F_p^N$.  It is easy 
to check, for example, that 
$6 = {\rm Tr}\,F_{p_1}^3 > {\rm Tr}\,F_{p_2}^3 = 3$; more generally 
${\rm Tr}\,F_p^N = \lambda_1^N + \cdots + \lambda_k^N$ when there are 
$k$ alternatives and $\lambda_i$ are the eigenvalues of $F_p$.

The thermodynamic formalism [9] provides a physical description of
symbolic dynamical systems.  Observe that
$$
{\rm Tr}\,F_p^N 
 = \lim_{T \to 0} \sum_{\sigma\in A^N} e^{-E_{f_p}(\sigma)/T}
 =: \lim_{T \to 0} Z_N[f_p,T],                                \eqno(1)
$$
where $A$ is the set of alternatives and, with the convention that 
$\sigma_{N+1} \equiv \sigma_1$,
$$
E_{f_p}(\sigma) := \sum_{i=1}^N 1-(F_p)_{\sigma_{i+1}\sigma_i}.
                                                              \eqno(2)
$$
$Z_N[f_p,T]$ is the {\sl partition function\/} for a statistical 
mechanics model on the lattice $\Z_N$ where the set of {\sl states\/} 
is $A$ and the {\sl energy\/} $E_{f_p}(\sigma)$ of a
{\sl configuration\/} $\sigma \in A^N$ is the sum of contributions 
from adjacent states:  0 if $\sigma_{i+1} \leftarrow \sigma_i$ in
$f_p$ and 1 otherwise.  The zero {\sl temperature\/} $T \to 0$ limit 
in Eq.\ 1 eliminates the contributions from all but the {\sl ground 
state\/} configurations, so the number of ground states (and hence 
$Z_N[f_p,0]$) is the same as the number of admissible $N$-periodic 
sequences.  From this perspective it is particularly natural to 
consider the {\sl free energy density\/} (the average energy per 
lattice site) or equivalently, the {\sl topological entropy\/} [10]:
$$
S[f_p] := \lim_{N\to\infty}{1\over N} \log Z_N[f_p,0].        \eqno(3)
$$
Inserting Eq.~1 into Eq.~3 we see that $S[f_p] = \log\Lambda_{f_p}$, 
where $\Lambda_{f_p}$ is the spectral radius of $F_p$, namely its 
largest eigenvalue.

The topological entropy measures the degree of mixing of the dynamical
system defined by $f_p$.  When the entropy is positive the dynamical
system is {\sl chaotic\/} and exhibits the familiar features of chaos:  
topological transitivity, sensitive dependence on initial conditions,
and a dense set of periodic points [11].  For the examples of Fig.~1,
we can compute $S[f_{p_1}] = 1$ (using logarithms in base 2) and 
$S[f_{p_2}] = 0$, which suggests that the presence of a cycle in $f_p$ 
makes the dynamical system chaotic.  This is true in general:

\noindent\Proposition.  {\sl The dynamical system defined by a 
complete directed graph has positive topological entropy iff the graph
contains a nontrivial cycle.}

\noindent\Proof.  If the directed graph has no nontrivial cycle there
is some ordering of the vertices for which the associated transition
matrix is upper triangular.  Hence all its eigenvalues are 1, so the
topological entropy vanishes.  Conversely, suppose there is a cycle of
length $l > 1$ in the directed graph.  Considering only those paths 
which lie entirely on the cycle, at each vertex of the cycle such a 
path may stay there or continue to the next vertex.  Starting from any 
vertex on the $l$-cycle, then, there are $2^N$ such paths of length 
$N$, which may require at most $l-1$ additional steps to close.  Thus
$S \ge \lim_{N\to\infty}(\log l2^N)/N = 1$.            \hfill\endproof

\moveright\secondstart\vtop to 0pt{\hsize=\halfwidth
$$
\epsfxsize=\halfwidth\epsfbox{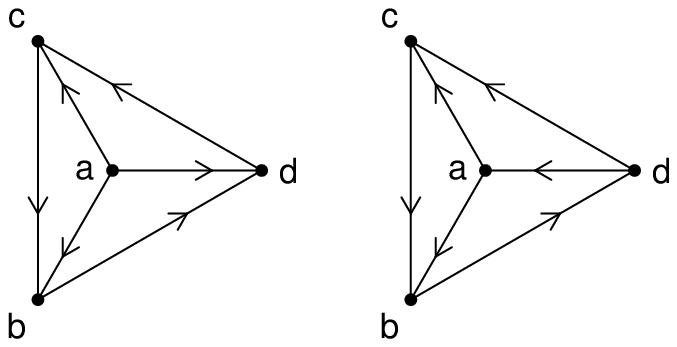}
$$
\vskip 0.25\baselineskip
\eightpoint{%
\noindent{\bf Fig.~2}.  The weak tournaments corresponding to 
majority rule on the profiles 
$p_3 = (b > c > a > d, c > d > a > b, d > b > c > a)$ and
$p_4 = (b > c > a > d, c > a > d > b, d > b > c > a)$.  On the left, 
$f_{p_3}$ contains a 3-cycle; on the right, $f_{p_4}$ contains a
4-cycle.
}}
\vskip -\baselineskip
\parshape=16
0pt \halfwidth
0pt \halfwidth
0pt \halfwidth
0pt \halfwidth
0pt \halfwidth
0pt \halfwidth
0pt \halfwidth
0pt \halfwidth
0pt \halfwidth
0pt \halfwidth
0pt \halfwidth
0pt \halfwidth
0pt \halfwidth
0pt \halfwidth
0pt \halfwidth
0pt \hsize
Since Arrow's theorem [2] guarantees the existence of cycles for any 
nondictatorial IIA voting rule and some profile on at least three 
alternatives, positivity of the topological entropy demonstrates the 
connection between Arrow's theorem, cycles and chaos hinted at by the 
`chaos' theorems in spatial voting models [12] as well as by Saari's 
suggestive analogies [3].  We therefore propose to use the topological 
entropy as a measure of the complexity of a profile/voting rule pair.  
It identifies the associated dynamical system as chaotic or nonchaotic 
and quantifies `how chaotic' the system is.  Consider the pair of 
profiles $p_3 = (b > c > a > d, c > d > a > b, d > b > c > a)$ and
$p_4 = (b > c > a > d, c > a > d > b, d > b > c > a)$ for three voters
and four alternatives.  Majority rule gives the weak tournaments shown 
in Fig.~2, both of which contain cycles and so define chaotic 
dynamical systems.  Notice that $f_{p_3}$ contains a 3-cycle while 
$f_{p_4}$ contains a 4-cycle; the entropies differ correspondingly 
[13]:
$$ 
S[f_{p_3}] = 1 \qquad {\rm and} \qquad S[f_{p_4}] \approx 1.260.
                                                              \eqno(4)
$$ 

Formulating the system as a statistical mechanics model focusses our
attention on the energy functional:  The energy of a configuration is 
defined (Eq.~2) by a matrix with entries $1 - (F_p)_{ab}$.  As the 
profile $p$ or the voting rule $f$ changes, this matrix can change.  
For a (probabilistic) ensemble of possible matrices the partition 
function (Eq.~1) defines a {\sl random matrix model\/} [14].

The first type of ensemble we consider is generated by a random 
distribution of profiles.  For a given voting rule, the statistical 
mechanics model defined by $Z_{N}[f_p,T]$ will be present in the 
ensemble with probability proportional to the number of profiles with 
the same image under $f$.  For example, again consider the situation 
of majority voting on three alternatives.  There are $(3!)^3$ profiles 
for three voters with strict preferences, out of which $2 \cdot 3!$ 
map to a weak tournament with a cycle as in $f_{p_1}$.  Thus for a 
random ensemble over these profiles, the average entropy for majority 
rule is $1 \cdot {1\over 18} + 0 \cdot {17\over 18} = {1\over 18} 
\approx 0.056$.  In the limit of an infinite (odd) number of voters, 
we can use a result originally obtained by Guilbaud [15], that the 
probability of a cycle is ${1\over4} - {3\over 2\pi}\arcsin{1\over3}$, 
to find that the average entropy for majority rule on three 
alternatives goes up to approximately 0.088.

Similarly, for four alternatives, there are $(4!)^3$ strict profiles 
for three voters, out of which 1632 map to a weak tournament with a 
3-cycle as in $f_{p_3}$ and 720 map to a weak tournament with a 
4-cycle as in $f_{p_4}$.  Weighting the entropies in Eq.~4 
accordingly, over this ensemble of profiles the average entropy for 
majority rule is approximately 0.184.  In the limit of an infinite 
(odd) number of voters, we can use a result of May and of Fishburn 
[16] that the probability of a Condorcet winner [13] is 
${1\over2} +{3\over\pi}\arcsin{1\over3}$, together with the result of 
Gehrlein and Fishburn [17] that the probability that there is no
nontrivial cycle is 
$$
{3\over8} + {6\over\pi^2}\int_0^{1/3} {\arccos[-x/(1-2x^2)] \over
                                       (1 - x^2)^{1/2}} {\rm d}x,
$$
to find that the average entropy for majority rule on four 
alternatives goes up to approximately 0.391.

We can also consider the same ensembles of profiles aggregated by 
other voting rules.  The Borda count [18], for example, assigns 
weights of $n-1$, $n-2$, $\ldots$, 0 to each voter's first, second, 
$\ldots$, last preferences, respectively, sums the weights of each 
alternative, and ranks the $n$ alternatives accordingly.  Since there 
is now the possibility of alternatives with equal ranks, even though 
the resulting weak tournament is transitive, it may still contain 
cycles [19].  For three voters the average entropy for the Borda count 
on three alternatives is 
$\log 3\cdot {1\over18} + 1 \cdot {1\over6} \approx 0.255$, while on 
four alternatives it is 
$\log 3\cdot {1\over18} + 1 \cdot {37\over96} \approx 0.473$.  For
comparison, we may use the Copeland method [20] to make the 
weak tournaments obtained by majority rule transitive:  assign each 
alternative a weight which is the number of incoming minus the number 
of outgoing edges and rank the alternatives accordingly.  For three 
voters the average entropy of the Copeland method on three 
alternatives is $\log 3 \cdot {1\over 18} \approx 0.088$, while on 
four alternatives it is 
$\log 3 \cdot {17\over 144} + 1 \cdot {5\over 96} \approx 0.239$, each
of which, although higher than for majority rule, is lower than the 
corresponding average entropy for the Borda count.

The second type of ensemble is generated by a random distribution of 
voting rules.  We consider, for example, a uniform distribution of
rules which satisfy IIA and have image in the set of strict 
tournaments.  ({\sl Strict tournaments\/} are weak tournaments with 
exactly one edge between every pair of vertices.)  Each such voting 
rule is defined by its images on the profiles restricted to all pairs 
of alternatives.  For a pair of alternatives there are $2^n$ 
possibilities for the restriction of a profile of $n$ strict 
preferences.  The voting rule maps each of these to an edge directed 
one of two ways between these alternatives in a strict tournament.  
Since voting rules are Pareto, the two unanimous restricted profiles 
have fixed images, but the remaining $2^n - 2$ may be mapped, 
independently, to either directed edge.  Thus, if there are $k$ 
alternatives, there are $2^{(2^n-2){k\choose 2}}$ possible IIA voting 
rules for $n$ voters.  Although this forms a huge ensemble of maps 
$f$, given a profile $p$, it is straighforward to determine with what 
probability each statistical mechanics model $Z_N[f_p,T]$ occurs in 
the ensemble.

For example, consider the profile 
$p_5 = (c > a > b > d, d > a > c > b, a > c > d > b)$.  Restricted to 
$\{a,b\}$ or to $\{b,c\}$ this profile is unanimous, so {\sl every\/}
voting rule, being Pareto, must map $p_5$ to a strict tournament with 
the edges $a \leftarrow b$ and $c \leftarrow b$.  The other pairwise 
restrictions, however, are not unanimous.  Since we are considering an
ensemble of IIA voting rules, this means that $a \leftarrow c$ (or
$c \leftarrow a$), $a \leftarrow d$ (or $d \leftarrow a$), 
$b \leftarrow d$ (or $d \leftarrow b$), and $c \leftarrow d$ 
(or $d \leftarrow c$) are independent events.  Furthermore, since the
voting rules in this ensemble map profiles to strict tournaments, each
of the 16 resulting possibilities has probability ${1\over 16}$.  No
further analysis of the ensemble is necessary; we can immediately
observe that with with probability ${3\over16}$ the strict tournament 
to which $p_5$ maps has a 4-cycle (like $f_{p_4}$), with probability
${5\over16}$ it has a 3-cycle (like $f_{p_3}$), and with probability
${1\over2}$ it is transitive.  Thus, using the entropies in Eq.~4, the 
average entropy for $p_5$ over the ensemble of strict IIA voting rules 
is approximately 
$1.260 \cdot {3\over16} + 1 \cdot {5\over16} \approx 0.549$.  We can
compare this to the average entropy for $p_3$ or $p_4$ over the same
ensemble.  Each of these profiles is only unanimous upon restriction
to $\{a,c\}$; the consequent probabilities for a 4-cycle, a 3-cycle, 
and transitivity are ${9\over32}$, ${11\over32}$, and ${3\over8}$, 
respectively, leading to a larger average entropy of approximately 
0.698.

This shows that the average entropy over an ensemble of voting rules
is a plausible measure of the cohesion or diversity in a society [21], 
as described by a profile:  When the number of pairs on which the 
profile is unanimous decreases, the average entropy increases.  
Furthermore, it is sensitive to {\sl which\/} pairs the voters rank 
consistently.  It is clear that for this IIA ensemble, unanimity on 
two disjoint pairs, \eg, $\{a,c\}$ and $\{b,d\}$, does not reduce the 
entropy from the value found for $p_3$ and $p_4$.

Taking a statistical mechanics approach to a general problem in 
social dynamics [1,2,6,12,21,22]---iterated preference 
aggregation---we have been led to the topological entropy as a 
quantitative measure of the complexity of profile/voting rule pairs.  
Unlike traditional approaches which have concentrated merely on the 
existence (or not) of cycles, use of this quantitative measure allows 
comparison between systems differing even in number of voters or 
alternatives.  Furthermore, we have constructed an annealed random 
matrix model for voting and considered ensembles corresponding to two 
natural social/political questions:  What level of complexity can we 
expect from a given voting rule?  How cohesive/diverse is the system 
relative to some collection of voting rules?  The first question has 
been addressed in previous work by evaluating the probabilities for 
the existence of a Condorcet winner or of a nontrivial cycle 
[15,16,17,23].  Such probabilities are inputs into our calculations of 
the average entropy for a given ensemble of profiles and voting rule.  
Our approach to the second question is a strong generalization of the 
usual analysis of the (non)existence of a cycle for a single voting 
rule.  Not only does the topological entropy provide for quantitative 
answers to these questions, it also makes precise the connection 
between the existence of cycles and chaos.  Increasing the number of 
autonomous agents or the number of alternatives increases the 
complexity of the system; chaos can be reduced or avoided only by 
changing [23] or severely restricting [2,6,21] the class of agent 
preferences and/or aggregation rules.

\medskip
\noindent{\bf Acknowledgements}
\nobreak

\nobreak
\noindent We gratefully acknowledge support from the John Deere 
Foundation and from DISA, and useful discussions with Mike Freedman, 
Des Johnston, Jean-Pierre Meyer and Melanie Quong.

\vfill\eject

\global\setbox1=\hbox{[00]\enspace}
\parindent=\wd1

\noindent{\bf References and Notes}
\bigskip

\parskip=0pt
\item{[1]}
M. J. A. N. de Caritat, Marquis de Condorcet,
{\it Essai sur l'application de l'analyse \`a la probabilit\'e des 
     d\'ecisions rendues \`a la pluralit\'e des voix\/}
(Paris:  {\it l'Imprim\`erie Royale\/} 1785).

\item{[2]}
K. J. Arrow,
{\sl Social Choice and Individual Values\/}
(New York:  Wiley 1951);\hfb
For proofs of Arrow's theorem from a perspective congenial to 
physicists, see\hfb
G. Chichilnisky,
``The topological equivalence of the Pareto condition and the
  existence of a dictator'',
\JME\ {\bf 9} (1982) 223--233;\hfb
Y. M. Baryshnikov,
``Unifying impossibility theorems:  a topological approach'',
\AAM\ {\bf 14} (1993) 404--415.

\item{[3]}
D. G. Saari,
``The ultimate of chaos resulting from weighted voting systems'',
\AAM\ {\bf 5} (1984) 286--308.

\item{[4]}
\brosl\ and M. W. Tilden,
``Living machines'',
\RAS\ {\bf 15} (1995) 143--169.

\item{[5]}
The adjective in `weak tournament' allows, in analogy with the `weak 
order' formalization of preference, the possibility that $f_p$ 
contain both $a \leftarrow b$ and $b \leftarrow a$.  That is, both
the individual preference orders in $p$ and their aggregate $f_p$ 
can include indifference between alternatives.  For an early use of 
directed graphs/tournaments to analyze voting, see
D. C. McGarvey,
``A theorem on the construction of voting paradoxes'',
\Ec\ {\bf 21} (1953) 608--610.

\item{[6]}
D. Black,
``On the rationale of group decision-making'',
\JPE\ {\bf 56} (1948) 23--34;\hfb
R. Farquharson,
{\sl Theory of Voting}
(New Haven:  Yale University Press 1969).

\item{[7]}
This condition was earlier called the {\sl postulate of relevancy\/} 
in E. V. Huntington,
``A paradox in the scoring of competing teams'',
\Sc\ {\bf 88} (1938) 287--288.

\item{[8]}
W. H. Gottshalk and G. A. Hedlund,
{\sl Topological Dynamics},
AMS colloquium publications, vol.\ 36
(Providence, RI:  AMS 1955).

\item{[9]}
D. Ruelle,
{\sl Thermodynamic Formalism:  The mathematical structures of 
     classical equilibrium statistical mechanics\/}
(Reading, MA:  Addison-Wesley 1978).

\item{[10]}
C. E. Shannon,
``A mathematical theory of communication'',
\BSTJ\ {\bf 27} (1948) 379--423; 623--656;\hfb
W. Parry,
``Intrinsic Markov chains'',
\TAMS\ {\bf 112} (1964) 55--66;\hfb
R. L. Adler, A. G. Konheim and M. H. McAndrew,
``Topological entropy'',
\TAMS\ {\bf 114} (1965) 309--319.

\item{[11]}
T. Li and J. Yorke,
``Period three implies chaos'',
\AMM\ {\bf 82} (1975) 985--992;\hfb
M. Misiurewicz,
``Horseshoes for continuous mappings of an interval'',
in C. Marchioro, ed.,
{\sl Dynamical Systems}, proceedings of the CIME session,
Bressanone, Italy, 19--27 June 1978
(Napoli, Italy:  {\it Liguori Editore\/} 1980) 125--135.

\item{[12]}
B. Ward,
``Majority rule and allocation'',
\JCR\ {\bf 5} (1961) 379--389;\hfb
\mckelvey,
``Intransitivities in multidimensional voting models and some 
  implications for agenda control'',
\JET\ {\bf 12} (1976) 472--482;\hfb
N. Schofield,
``Instabilities of simple dynamic games'',
\RES\ {\bf 40} (1978) 575--594;\hfb
L. Cohen, 
``Cyclic sets in multidimensional voting models'',
\JET\ {\bf 20} (1979) 1--12;\hfb
\mckelvey,
``General conditions for global intransitivities in formal voting 
  models'',
\Ec\ {\bf 47} (1979) 1085--1112;\hfb
E. C. Browne, P. A. James and M. A. Miller,
``Simulation of global and local intransitivities in a simple 
  voting game under majority rule'',
\BS\ {\bf 36} (1991) 148--156;\hfb
D. Richards,
``Intransitivities in multidimensional spatial voting:  period 
  three implies chaos'',
\SCW\ {\bf 11} (1994) 109--119.

\item{[13]}
We must also note, however, that the entropy does not distinguish 
between the cycle in $f_{p_3}$ being a top cycle and being a bottom 
cycle, as it would be were all the edges from $a$ reversed, making 
$a$ the Condorcet winner (a pairwise unbeatable alternative).

\item{[14]}
E. P. Wigner,
``Characteristic vectors of bordered matrices with infinite 
  dimensions'',
\AnM\ {\bf 62} (1955) 548--564;\hfb
M. L. Mehta and M. Gaudin,
``On the density of eigenvalues of a random matrix'',
\NP\ {\bf 18} (1960) 420--427;\hfb
F. J. Dyson, 
``Statistical theory of the energy levels of complex systems.  
  I, II, III'',
\JMP\ {\bf 3} (1962) 140--175.

\item{[15]}
G. Th.\ Guilbaud,
``{\it Les th\'eories de l'int\'er\^et g\'en\'eral et le probl\`eme
       logique de l'aggr\'ega\-tion}'',
\EA\ {\bf 5} (1952) 501--584.

\item{[16]}
R. M. May,
``Some mathematical remarks on the paradox of voting'',
\BS\ {\bf 16} (1971) 143--151;\hfb
P. C. Fishburn,
``A proof of May's theorem $P(m,4) = 2 P(m,3)$'',
\BS\ {\bf 18} (1973) 212.

\item{[17]}
W. V. Gehrlein and P. C. Fishburn,
``Probabilities of election outcomes for large electorates'',
\JET\ {\bf 19} (1978) 38--49.

\item{[18]}
J.-C. de Borda,
``{\it M\'emoire sur les \'elections au scrutin}'',
\MARS\ (1781) 657--665.

\item{[19]}
The Borda count is not an IIA voting rule, so Arrow's theorem [2]
does not apply.  Nevertheless, alternatives may be equally ranked
and, if we define the dynamics by choosing an equally ranked new
alternative over the {\it status quo}, the system will then 
contain cycles.

\item{[20]}
A. H. Copeland,
``A `reasonable' social welfare function'',
mimeographed notes,
University of Michigan seminar on applications of mathematics to
 the social sciences
(November 1951);\hfb
referred to in 
L. A. Goodman,
``On methods of amalgamation'',
in R. M. Thrall, C. H. Coombs and R. L. Davis, eds.,
{\sl Decision Processes\/}
(New York:  Wiley 1954) 39--48.

\item{[21]}
G. Chichilnisky,
``Social diversity, arbitrage, and gains from trade:  a unified
  perspective on resource allocation'',
\AER\ {\bf 84} (1994) 427--434.

\item{[22]}
\dajm,
``Towards the global:  complexity, topology and chaos in modelling,
  simulation and computation'',
chao-dyn/9710005,
\IJCS, Article [123],
http://dynamics.bu.edu/InterJournal/.

\item{[23]}
B. Jones, B. Radcliff, C. Taber and R. Timpone,
``Condorcet winners and the paradox of voting:  probability 
  calculations for weak preference orders'',
\APSR\ {\bf 89} (1995) 137--144.

\bye